\documentclass[conference]{IEEEtran}
\usepackage{epsfig}
\usepackage{graphicx}
\usepackage{subfigure}
\def\v{{\mathbf v}}

\def\x{{\mathbf x}}
\def\y{{\mathbf y}}
\def\zt{{\mathbf{z}_{\mathrm{train}}}}
\def\ztt{{\mathbf{z}^*_{\mathrm{train}}}}
\def\yt{{\mathbf{y}_{\mathrm{train}}}}
\def\xt{{\mathbf{x}_{\mathrm{train}}}}
\def\xf{{\mathbf{x}_{\mathrm{frequency}}}}
\def\yf{{\mathbf{y}_{\mathrm{frequency}}}}
\def\yi{{\mathbf{y}_{\mathrm{impulse}}}}
\def\xi{{\mathbf{x}_{\mathrm{impulse}}}}

\def\yti{{\left(\mathbf{y}_{\mathrm{train}}\right)_i}}

\def\zti{{\left(\mathbf{z}_{\mathrm{train}}\right)_i}}
\def\xd{{\mathbf{x}_{\mathrm{data}}}}
\def\yd{{\mathbf{y}_{\mathrm{data}}}}
\def\hg{{\mathbf{h}_{\mathrm{gaussian}}}}
\def\hc{{\mathbf{h}_{\mathrm{constant}}}}
\def\T{{T_{\mathrm{threshold}}}}
\def\SNRf{{\mathrm{SNR}_{\mathrm{frequency}}}}
\def\z{{\mathbf z}}

\def\h{{\mathbf h}}

\usepackage{epsf}
\usepackage{ifthen}

\newcommand{\thd}[1]
{\ifthenelse {\equal{#1}{1}}
	{{#1}$^{\mathrm{st}}$} 
	{{\ifthenelse{\equal{#1}{2}}{{#1}$^{\mathrm{nd}}$}{{#1}$^{\mathrm{th}}$}}}}

\newcommand{\SNR}{\mathrm{SNR}}
\newcommand{\mmse}{\mathrm{mmse}}

\newtheorem{Theorem}{Theorem}

\usepackage{amsmath}

\newtheorem {Corollary}  [Theorem]    {Corollary}

\begin{document}
\title{TRAINING OVER SPARSE MULTIPATH CHANNELS IN THE LOW SNR REGIME}
\author{\IEEEauthorblockN{Elchanan Zwecher}
\IEEEauthorblockA{School of Engineering and\\Computer Science\\
Jerusalem 
Israel\\
Email: elchanzw@cs.huji.ac.il}
\and
\IEEEauthorblockN{Dana Porrat}
\IEEEauthorblockA{School of Engineering and\\Computer Science\\
Jerusalem 
Israel\\
Email: dana.porrat@huji.ac.il}
}
\maketitle
\begin{abstract}
Training over sparse multipath noisy channels is explored. The energy allocation and the optimal shape of training signals that enable communications over unknown channels are characterized as a function of the channels' statistics. The performance of training is evaluated by the reduction of the mean square error of the channel estimate and by the decrease in the \emph{penalty term}- the mutual information reduction due to the uncertainty of the channel. The performance of low dimensional training signal is compared to the performance of a full dimensional one. Especially, The trade-off between the number of required measurements (signal dimensions) and the energy allocation is calculated, and it is proven that if the signal to noise ratio of the received training signal is low, reducing the number of channel measurements using compressed sensing is efficient in the sense of energy consumption. 
\end{abstract}
\section{Introduction}\label{sec:introduction}
Channel statistics determine the incoherent achievable rate~\cite{zheng_OnChannel,porrat_2007Channel,zwecher_Spreading}. If we transmit $\x$ over a noisy random LTI channel denoted by the random impulse response $\h$, and white Gaussian noise $\z$ is added such that the received signal is $\y=\h\ast\x+\z$, where $\ast$ denotes convolution, then the mutual information between $\y$ and $\x$ obeys
\begin{eqnarray}
I(\y;\x)&=&I(\y;\x,\h)-I(\y;\h|\x)\label{eq:penalty_term0}\\
&\geq&I(\y;\x|\h)-I(\y;\h|\x)\nonumber
\end{eqnarray}
The second form term of~(\ref{eq:penalty_term0}) 
$I(\y;\h|\x)$
is the \emph{penalty term} due to the uncertainty of the channel $\h$. The mutual information between $\y$ and $\x$ is lower bounded by coherent rate minus the penalty term. The penalty term is a function of the statistics of $\h$, the 'richer' is the statistics of $\h$ i.e.~the bigger the entropy of $\h$, the higher the penalty term.\\ 
The statistics of the channel affect its entropy, but what is their effect on the best way to train the system? 
This paper is concentrated on training over the sparse multipath channel. 
%
This channel can be considered as a collection of narrowband eigenchannels, with no interference between them. Each eigenchannel amplifies the transmitted data by a gain and as a result of channel sparsity, there is a dependence between the gains of the eigenchannels, dependence that causes the low uncertainty of the channel. \\
The performance of training can be evaluated from two perspectives: the minimum mean square error (MMSE) in estimating the channel and the reduction in the penalty term.\\
By compressed sensing, one can divide the signal in the frequency domain to a data part and a training part. \\
Recovering sparse vectors in noisy environments using thresholding is discussed in~\cite{donoho}. The idea of recovering sparse vectors after compressing them is introduced in~\cite{candes_2004_a,candes_2004_b} and this ability was extended to the noisy case~\cite{candes_2005}. Recent works are concentrated on the ability of exact pattern recovery~\cite{aeron_2009,fletcher_2009,wang_2010}, i.e.~the ability to detect almost always \textbf{all} the non-zero entries of the vector $\h$ which represents the channel. \cite{guo_2009} discusses compressed sensing of vectors and~\cite{bajwa_2009_a} discuss compressed sensing of channels in the finite SNR regime.\\
A connection between information theory and compressed sensing is introduced in~\cite{sarvotham_2006}. This work bounds the number of required measurements (the rows' rank of the the compressing matrix) needed to reduce the mean square error of the compressed random vector $\v$ to a value of $\eta\in\mathcal{R}^+$ in a noisy environment:
\begin{equation}
m\geq \frac{\mathcal{R}_{\v}(\eta)}{\frac{1}{2}\log(1+\SNR)}\label{eq:baron_bound}
\end{equation}
where $\mathcal{R}_{\v}(\eta)$ is the rate distortion function of $\v$ at the point $\eta$ and $\frac{1}{2}\log(1+\SNR)$ is the capacity of an $\mathrm{AWGN}$ channel. 
From~(\ref{eq:baron_bound}) the total energy of compressing/training is lower bounded by
 \begin{equation}
 m\SNR\geq 2\mathcal{R}_{\v}(\eta)\label{eq:baron_energy}
 \end{equation}
 However, it is not clear whether the bounds~(\ref{eq:baron_bound}) and~(\ref{eq:baron_energy}) are achievable.\\
\textbf{Our Contribution:}
In accordance with the physical characteristics of multipath channels in the wideband limit we assume that the sparsity of the channel tends to zero and that the channel remains constant during short coherence periods. Unlike papers that discuss finite or high SNR~\cite{guo_2009,aeron_2009}, this paper is concentrated in training in the low SNR regime, where sparsity enables recovery of the channel. We design a signal composed of a data part and a training part such that the output can be separated into two parts that do not interfere, and training uses as small a subspace as possible so the data space is maximized.\\
\cite{fletcher_2009,wang_2010} look for exact pattern recovery, and their results do not achieve the lower bounds~(\ref{eq:baron_bound})~(\ref{eq:baron_energy}). We show that in the low SNR regime, if one is satisfied with \textbf{almost} perfect channel recovery then using techniques of compressed sensing the lower bound on the number of require measurements~(\ref{eq:baron_bound}) is achievable while using minimum training energy~(\ref{eq:baron_energy}), as long as the training signal is composed of enough harmonic vectors and channel measurements are done in the frequency domain. In addition we evaluate the effect of training on the penalty term.\\
A comparison of the required training energy and number of channel measurements between this paper and~\cite{fletcher_2009} is presented in Table~\ref{fig:compare}.\\  
\begin{table}\label{fig:compare}
\begin{tabular}{| p{1cm}| p{1.5cm}|p{2.5cm}|p{2cm}|}
\hline
&reconstruction error&number of measurements&total training energy\\
\hline
Fletcher et al.\cite{fletcher_2009}&0&$\frac{8}{\SNR}\times(1+\SNR)\times L\log(k_c-L)$&$\approx 8L\log(k_c-L)$\\
\hline
This paper&$o(1)$&$\omega\left(L\log \frac{k_c}{L}\right)$&$2\left(L\log \frac{k_c}{L}\right)$\\
\hline
\end{tabular}
\caption{A comparison between the number of required measurements and total training/compressing energy in order to reduce the mean square error from Fletcher et al.~\cite{fletcher_2009} and from this paper. The random channel $\h$ is unit norm, its length is $k_c$ and the number of non-zero entries is $L$. Unlike~~\cite{fletcher_2009} that search for perfect recovery, we allow negligible mean square error that enables the reduction of the training energy by a factor of at least four.}
\end{table} 
\section{Channel model and training signals}\label{sec:channel_signal_model}
The model assumes that the channel remains constant during a period of $t_c$, and the maximal delay is $t_d$, where $t_d$ is significantly smaller than $t_c$ and both are constants independent of the bandwidth. As the bandwidth increases the number of delayed reflections of the transmitted signal also grows, however the growth is sublinearly with the bandwidth. The receiver gets the signal with additive white Gaussian noise independent of the transmitted signal and the channel. After discretizing, the channel can be represented by the vector $\h$ of length $k_c=w\times t_c$, where $w$ is the bandwidth, the last $w(t_c-t_d)$ entries are zero and only the first $k_d=w\times t_d$ entries may differ from zero. 
\subsection{Channel model}\label{sec:channel_model}
The statistics of each path delay are as follows: each of the first $k_d$ entries of $\h$ is an active path in probability $\frac{\mathcal{L}(w)}{k_d}$ independent of the other entries, and $\lim_{w\to\infty}\frac{\mathcal{L}(w)}{k_d}=0$ so the channel becomes sparser as the bandwidth increases.
The statistics of the amplitudes are also independent and denoted by the probability density function $\mathcal{P}(\cdot)$.
 We assume that $E\left[\h\right]=0$ and $E\left\|\h\right\|_2^2=1$ such that each amplitude of an active path has a zero mean and variance $\frac{1}{\mathcal{L}(w)}$.
 Since $k_d$ is significantly smaller than $k_c$, we approximate the result of the LTI channel by a cyclic convolution and get the received training signal 
 $\yt=\sqrt{\SNR}\xt\ast \h+\zt$ 
where $\ast$ denotes the cyclic convolution, $\zt$ is additive white Gaussian noise and $\SNR$ is the signal to noise ratio.\\
Let $\xd$ be the data part of the transmitted signal. If we transmit and train concurrently then $\y=\left(\sqrt{\SNR}\xt+\xd\right)\ast \h+\z$. The realization of the channel depends on the pdf of the path gains $\mathcal{P}$. We pay special attention to the following cases: \textbf{(1)} the statistics of the path gains are Gaussian and \textbf{(2)} the active path gains equal $\frac{1}{\sqrt{\mathcal{L}(w)}}$ in probability $\frac{1}{2}$ and $-\frac{1}{\sqrt{\mathcal{L}(w)}}$ in probability $\frac{1}{2}$ so the absolute value of the gains is constant. In the Gaussian case we replace $\h$ by $\hg$ and in the constant (absolute value) case we replace $\h$ by $\hc$. \\ 
Hence, in each general formula contains $\h$, a subscript may be used to denote the type of che channel (constant or gaussian). 
\subsection{Training signals}
We introduce two training signals: 
\begin{itemize}
 \item $\xi$ that uses all the transmission space (i.e.~all the eigenchannels get training energy).
\item $\xf$ that uses only part of the transmission space.
\end{itemize}
\subsubsection{Impulse probing}
Impulse probing means sending a pulse over the channel $\h$ to get its noisy impulse response. The impulse training signal is:
\begin{equation}
 \left(\xi\right)_i=\left\{\begin{array}{cc}\sqrt{k_c}&i=1\\0& \mathrm{otherwise}\end{array}\right.\nonumber
\end{equation}
Training with $\xt=\xi$ we get the received training signal:
\begin{equation}
\yi=\sqrt{\SNR}\sqrt{k_c}\h+\zt\label{eq:training_impulse}
\end{equation}
\subsubsection{Training in the frequency domain}\label{sec:frequency_domain}
This type of training uses fewer measurements and enables to divide the band to a data and training band. The eigenvectors of the LTI channel are the harmonic vectors. The $i$'th $k_c$-length harmonic vector $\mathbf{f}^{(i)}$  at the $k$'th position obeys:
\[
\mathbf{f}^{(i)}_k=\frac{1}{\sqrt{k_c}}e^\frac{2\pi ji(k-1)}{k_c}\;i=1,2,...,k_c\nonumber
\] 
The training signal is chosen randomly in the following way: Let $\mathcal{Q}$ be a $m$-random subset of $\left\{1,2,...,k_c\right\}$. The training signal $\xf$ is:
\begin{equation}
\xf=\sqrt{\frac{k_c}{m}}\sum_{i\in\mathcal{Q}}\mathbf{f}^{(i)}\label{eq:xf}
\end{equation}
Training using $\xf$  we get:
\begin{equation}
\yt=\sqrt{\SNR}\xf\ast\h+\zt\label{eq:frequency_training}
\end{equation}
Let $i_1,i_2,...,i_m$ be the elements of $\mathcal{Q}$ and let $\lambda_i,\; i=1,2,..,i_{m}$ be the eigenvalues of the cyclic convolution represented by $\h$, i.e. $\h\ast\mathbf{f}^{(i)}=\lambda_i\mathbf{f}^{(i)}$. 
Obviously:
\begin{eqnarray}
\yt&=&\sqrt{\SNR}\xf\ast\h+\zt\\
&=&\sum_{i_j\in\mathcal{Q}}\sqrt{\SNR}\sqrt{\frac{k_c}{m}}\lambda_{i_j}\mathbf{f}^{(i_j)}+\zt\label{eq:received_frequency}
\end{eqnarray}
Let $F$ be a matrix whose rows are the harmonic vectors corresponding to $\mathcal{Q}$. 
Projecting $\yt$ onto $F$ and using the orthogonality of harmonic vectors we get the vector $\yf$:
\begin{equation}
 \yf=F\yt=\sqrt{\SNR}\sqrt{\frac{k_c}{m}}\left(\begin{array}{c}\lambda_{i_1}\\\lambda_{i_2}\\\cdots\\\lambda_{i_m}\end{array}\right)+\ztt\label{eq:yf2}
\end{equation}
where $\ztt$ is white Gaussian noise with unit norm.
The convolution~(\ref{eq:frequency_training}) is equivalent to projecting $\h$ onto the compressing matrix $F$. 
Since $E\left[\lambda_i^2\right]=1$, $\SNRf$, the signal to noise ratio of $\yf$ is
\begin{equation}
\SNRf=\frac{k_c}{m}\SNR\label{eq:SNRf}
\end{equation}
We can now compare training by impulse probing to compressed training in the frequency domain: in both cases~(\ref{eq:training_impulse}) and~(\ref{eq:frequency_training}) the total energy of training is
$
\SNR\left\|\xi\right\|_2^2=\SNR\left\|\xf\right\|_2^2=\SNR k_c\nonumber
$
However, $\yi$, the received training signal of $\xi$ is $k_c$-length with signal to noise ration $\SNR$ while $\yf$ is $m$-length with signal to noise ratio $\frac{k_c}{m}\SNR$.
\section{Performance of Ttraining}\label{sec:performance_of_training}



\subsection{Minimum mean square error}
The minimum mean square error of $\h$ given $\yt$ is $E\left[\left\|\h-E\left[\h|\yt\right]\right\|_2^2\right]$. Let $\yt(\SNR)=\sqrt{\SNR}\xt\ast \h+\zt$. The minimum mean square error of $\h$ given $\yt$ as a function of $\SNR$ is
\begin{equation}
\mmse(\SNR)=E\left[\left\|\h-E\left[\h|\yt(\SNR)\right]\right\|_2^2\right]\label{eq:mmse_function}
\end{equation}
Obviously, the higher the SNR the smaller the minimum mean square error so~(\ref{eq:mmse_function}) monotonically decreases. Later (in Theorem~\ref{th:1}) we see that the curve of the function behaves as a decreasing step function.
\subsection{Penalty term and rate distortion function} \label{sec:penalty_term_and_rate_distortion_function}
An alternative way to quantify the performance of training is to evaluate the uncertainty of the channel after training. We introduce two such similar criterias: penalty term and rate distortion function. Section~\ref{sec:Training_by_Impulse_Probing} shows that using a low training energy, minimum mean square error is not reduced although the penalty term and the rate distortion function are strongly affected.  
\subsubsection{Rate distortion function}

Let $\eta_0$ be a small (negligible) positive number. The rate distortion function $\mathcal{R}_{\h}(\eta_0)$ quantifies the amount of information required to almost perfectly recover the channel. If we have already trained the system, the remaining amount of information required to recover the channel is reduced
$\mathcal{R}_{\h|\yt}(\eta_0)<\mathcal{R}_{\h}(\eta_0)$.
The rate distortion function without training can be approximated by
\begin{equation}
\mathcal{R}_{\h}(\eta_0)\approx (1+o(1))k_d\times \mathcal{H}_b\left(\frac{\mathcal{L}(w)}{k_d}\right)\label{eq:rdf_eta0}
\end{equation}
when $\mathcal{H}_b(\cdot)$ is the binary entropy function. (\ref{eq:rdf_eta0}) is justified because the information required for an approximate recovery of $\h$ is a discrete $k_d$-length vector which contains the information on the path delays plus $\mathcal{L}(w)$ variables that contain data about the path gains. However, the required information on the path gains is negligible relative to the required information on the path delays (see~\cite{porrat_2007Channel}). 
Let $\mathcal{R}_{\h}^{(\eta_0)}(\SNR)$ be the rate distortion function after training as a function of $\SNR$.
\begin{eqnarray}
\mathcal{R}_{\h}^{(\eta_0)}(\SNR)&=&\mathcal{R}_{\h|\yt(\SNR)}(\eta_0)\nonumber\\
&\leq& (1+o(1))k_d\times \mathcal{H}_b\left(\frac{\mathcal{L}(w)}{k_d}\right)\label{eq:rdf_eta0_after_training}\\
&-&I\left(\yt(\SNR);\h\right)\nonumber
\end{eqnarray}
A comparison between the rate distortion function and the minimum mean square error after training is possible by comparing Figure~\subref{fig:subfig1} to Figure~\subref{fig:subfig2}. 
\subsubsection{Penalty term}
The penalty term, the reduction in mutual information due to the uncertainty of channel, is the mutual information between the received data signal $\yd$ and the channel $\h$. Under resonable assumptions on the data and training signals, the penalty term equals $I(\yd;\h|\yt(\SNR))$ and is upper bounded by~(\ref{eq:rdf_eta0_after_training}):
\begin{eqnarray}
&&I\left(\yd;\h|\xd,\yt(\SNR)\right)\label{eq:channel_entropy}\\
&&\leq (1+o(1))k_d\times \mathcal{H}_b\left(\frac{\mathcal{L}(w)}{k_d}\right)
-I\left(\yt;\h\right)\nonumber
\end{eqnarray}
\subsection{Optimization}
Optimization of the training is done over the number of required measurements and the energy consumption, so we want to minimize the energy of $\xt$ and when training in the frequency domain also the number of harmonic vectors composing $\xf$. From~\cite{sarvotham_2006} we know that the number of required measurements $m$ for negligible minimum mean square error $\eta_0$ is lower bounded by 
$m\geq \frac{\mathcal{R}_{\h}(\eta_0)}{\frac{1}{2}\log(1+\SNR)}$
so the required energy is lower bounded by $
m\SNR\geq \SNR\frac{\mathcal{R}_{\h}(\eta_0)}{\frac{1}{2}\log(1+\SNR)}\geq 2\mathcal{R}_{\h}(\eta_0)\label{eq:almost_perfect}$.
The following section show that these bounds are achievable.
\section{Training by impulse probing}\label{sec:Training_by_Impulse_Probing}
\subsection{Minimum Mmean square error and rate distortion function of $\hc$}
\subsubsection{Minimum mean square error}\label{sec:mean_square_error}
Let $\epsilon$ be a positive number as small as we wish and let
\begin{equation}
\SNR_0= \frac{2k_d\mathcal{H}_b\left(\frac{\mathcal{L}(w)}{k_d}\right)}{k_c}\label{eq:SNR0}
\end{equation}
The following theorem shows the effect of the training energy on the mean square error of channel recovery:
\begin{Theorem}\label{th:1}
If the total training energy is at least 
\begin{equation}
(1+\epsilon)k_c\SNR_0\label{eq:training_energy}
\end{equation}
then the minimum mean square error of $\hc$ is o(1) in the wideband limit. On the other hand, if the total training energy is less than~(\ref{eq:training_energy}) then the asymptotic mean square error of $\hc$ is $1-o(1)$.
\end{Theorem}
\textbf{Sketch of proof:} Let $\T=\sqrt{\frac{k_c}{\mathcal{L}(w)}\SNR_0}$. The proof is based on he fact that only $o\left(\mathcal{L}(w)\right)$ noise terms are high such that $\left|\zti\right|\geq\T$ but as long as the training energy is higher than~(\ref{eq:training_energy}), almost every $\yti$ corresponding to an active path obeys $\left|\yti\right|\geq\T$ so recovery is almost perfect and negligible minimum mean square error is achievable. \\
On the other hand, if we use a little less training energy than~(\ref{eq:training_energy}), then the $\left|\yti\right|$'s corresponding to active paths do not achieve the threshold $\T$, and there are much more than $\mathcal{L}(w)$ noise terms that are bigger than most of the $\left|\yti\right|$'s whose origins are active paths so random noise terms are more likely to look like active paths than the actual ones. As a result, \textbf{any} estimator cannot decide whether the origin of $\yti$ is an active path or a noise term and the estimation completely fails.\\
\textbf{Interpertation of Theorem~\ref{th:1}:} This theorem in fact shows that the required training energy for almost perfect channel recovery asymptotically achieves the lower bound~(\ref{eq:baron_energy}). 
To see this remember from~(\ref{eq:baron_energy}) and~(\ref{eq:rdf_eta0}) that the required training energy to recover the channel is lower bounded by
\begin{equation}
2\mathcal{R}_{\h}(\eta_0)=2(1+o(1))k_d\times \mathcal{H}_b\left(\frac{\mathcal{L}(w)}{k_d}\right)\label{eq:training_recovery}
\end{equation}
Combining~(\ref{eq:SNR0}),~(\ref{eq:training_energy}) and~(\ref{eq:training_recovery}),  the training energy of Theorem~\ref{th:1} achieves the lower bound on training energy~(\ref{eq:baron_energy}), because $\epsilon$ in~(\ref{eq:training_energy}) is as small as we wish.
The mean square error as a function of $\SNR$ behaves approximately as a step function, because the mean square error of $\h$ is $1-o(1)$ if $\SNR\leq (1-\epsilon)\SNR_0$ and $o(1)$ if $\SNR\geq(1+\epsilon)\SNR_0$. The reduction in the minimum mean square error occurs in the interval $\left[(1-\epsilon)\SNR_0,(1+\epsilon)\\SNR_0\right]$, which is as small as we wish. 
\subsection{Mean square error, penalty term and rate distortion function of $\hc$}
Although training with limited energy may be inefficient in the sense that it does not reduce the mean square error, it does affect the penalty term.
Using the I-MMSE connection we conclude from Theorem~\ref{th:1}:
\begin{Corollary}\label{th:col1}
The penalty term of $\hc$ after training~(\ref{eq:channel_entropy}) is upper bounded by:
\begin{eqnarray}
&&I\left(\yd;\hc|\xd,\yt(\SNR)\right)\nonumber\\
&&\leq(1+o(1))k_d\times \mathcal{H}_b\left(\frac{\mathcal{L}(w)}{k_d}\right)-I\left(\yt;\h\right)\nonumber\\
&&=(1+o(1))k_d\mathcal{H}_b\left(\frac{\mathcal{L}(w)}{k_d}\right)-\frac{1}{2}\int_{s=0}^{s=\SNR}\mmse(s)ds\nonumber\\
&&=(1+o(1))k_d\mathcal{H}_b\left(\frac{\mathcal{L}(w)}{k_d}\right)-\nonumber\\
&&\left\{\begin{array}{ll} \frac{1}{2}k_c\SNR&\SNR\leq (1-\epsilon)\SNR_0\\(1-o(1))k_d\mathcal{H}_b\left(\frac{\mathcal{L}(w)}{k_d}\right)&\SNR\geq (1+\epsilon)\SNR_0\end{array}\right.\label{eq:mmse}
\end{eqnarray}
%
 \end{Corollary}
\textbf{Interpretation:}
Since the mean square error is a step function of SNR, the mutual information between $\yt$ and $\hc$ increases linearly when recovery fails and remain constant when recovery is almost perfect. As a result the penalty term decreases linearly to a negligible value.\\ 
\subsection{Mean square error, penalty term and rate distortion function of $\hg$}
The ability to detect a path delay depends on its gain's impulsivity. Like in Theorem~\ref{th:1}, training can detect with high probability the delays of active paths of $\hg$ as long as 
$\left|\yti\right|=\left|\sqrt{\SNR}\sqrt{k_c}h_i+\zti\right|\geq \T$
We begin with a theorem summarizing the results of estimating $\hg$ and then compare them to the $\hc$ case.
\begin{Theorem}\label{th:2}
Let $\mathcal{Q}(\cdot)$ be the cummultive density function of normal random variable. The minimum mean square error of $\hg$ as a function of SNR obeys:
\begin{equation}
\mmse(\SNR)\approx\int_{s=-\sqrt{\frac{\SNR_0}{\SNR}}}^{s=\sqrt{\frac{\SNR_0}{\SNR}}}s^2\mathcal{Q}(s)ds\label{eq:mmse_hg}
\end{equation}
\end{Theorem}
The proof is omitted.
%

\begin{figure}\label{fig:2}
\centering
\subfigure[Minimum mean square error of $\hc$ (upper curve) vs. $\hg$ (lower curve). Using very low $\SNR$, training over $\hg$ is more efficient.]{
\includegraphics[width=100pt]{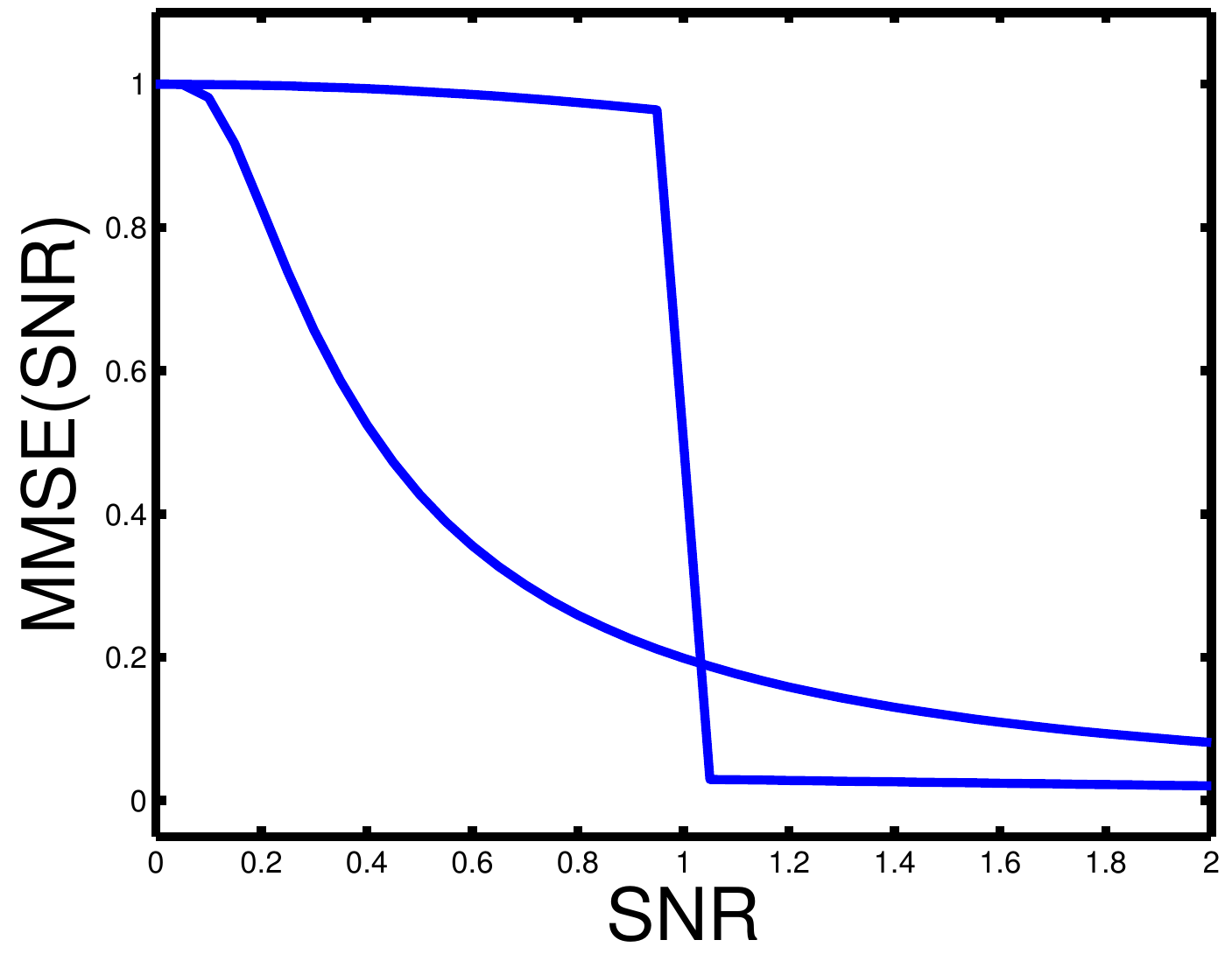}
\label{fig:subfig1}
}
\subfigure[The ratio between the rate distortion function after training $\mathcal{R}_{\h}^{(\eta_0)}(\SNR)$ and the initial rate distortion function $\mathcal{R}_{\h}(\eta_0)$ of $\hc$ (lower curve) vs. $\hg$ (upper curve). In any SNR, training $\hc$ yields better results.]{
\includegraphics[width=100pt]{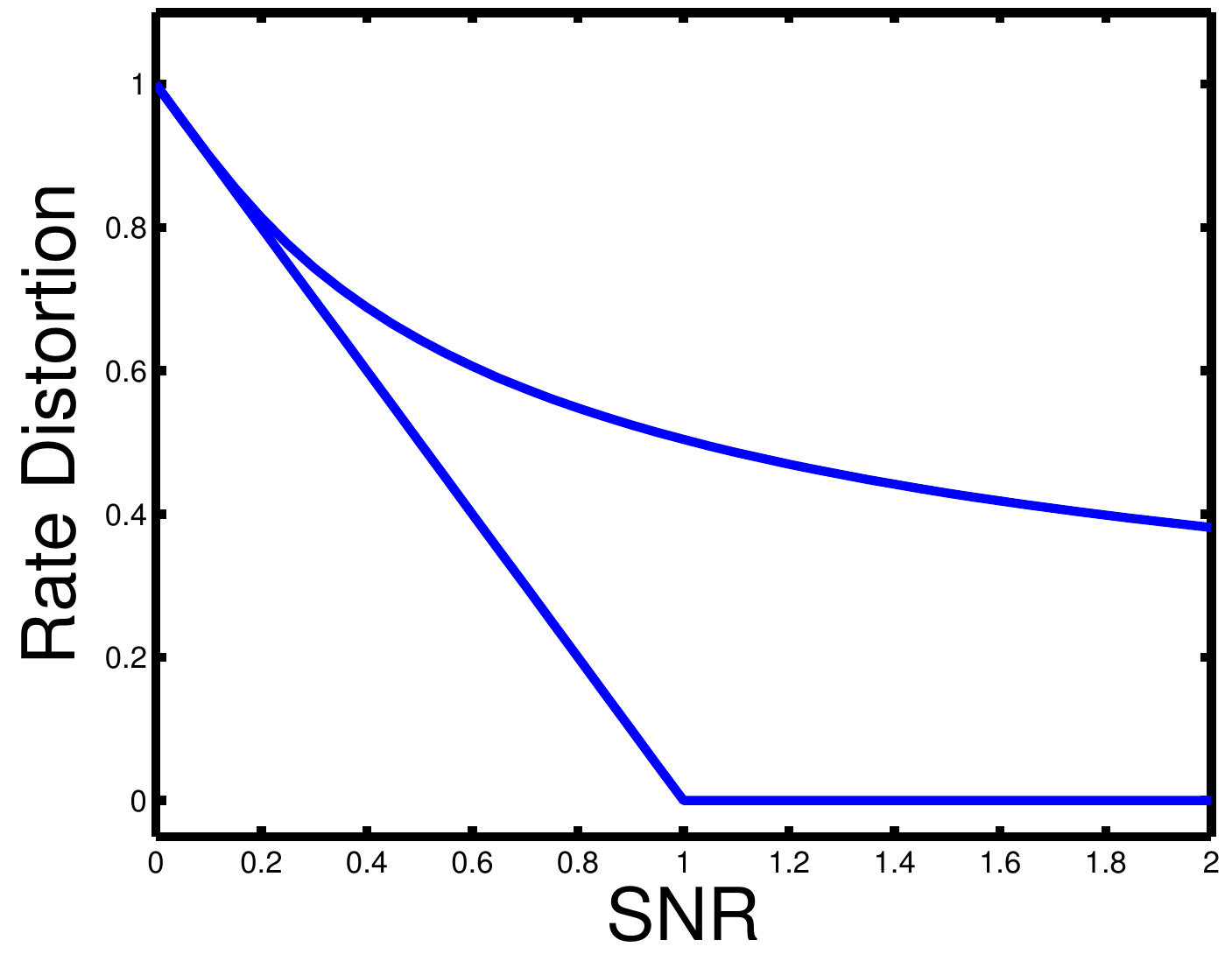}
\label{fig:subfig2}
}
\end{figure}

Using the I-MMSE connection, we get the following corollary regarding the penalty term of the estimate of $\hg$
\begin{Corollary}\label{th:col2}
The penalty term~(\ref{eq:channel_entropy}) of $\hg$ is upper bounded by:
\begin{eqnarray}
&&I\left(\yd;\hg|\xd,\yt(\SNR)\right)\\
&&=(1+o(1))k_d\times \mathcal{H}_b\left(\frac{\mathcal{L}(w)}{k_d}\right)\label{eq:col2}-\frac{1}{2}\int_{s=0}^{s=\SNR}\mmse(s)ds\nonumber
\end{eqnarray}
\end{Corollary}
The penalty term~(\ref{th:col2}) does not decrease linearly as in the $\hc$ case, but in a strictly convex manner, see Figure~\subref{fig:subfig2}.  
\textbf{Interpretation of Theorem~\ref{th:2} and Corollary~\ref{th:col2}}
\begin{enumerate}
\item
The performance of training in terms of minimum mean square error of $\hg$ as $\SNR\leq (1-\epsilon)\SNR_0$ is better than training over $\hc$ (compare Theorem~\ref{th:1} to Theorem~\ref{th:2}). However, as $\SNR\geq (1+\epsilon)\SNR_0$ training $\hc$ yields better results. Anyway, in terms of penalty term training over $\hc$ is more efficient at any SNR, see Figure~\subref{fig:subfig1} and Figure~\subref{fig:subfig2}.
\item
The performance of training $\hg$ depends on the impulsivity of the path gains, and is not due to their uncertainty. If the path gains where Gaussian and known, the asymptotic results where identical to results over $\hg$ although in the $\hg$ model the amplitudes are not known.
\item
The mean square error, unlike the penalty term, is very sensitive to the extreme noise values. Since modeling physical noise as white Gaussian relates to the average case, it is interesting to measure the behavior of the extreme case of the physical noise in multipath channels. Note that the extreme case 'captures' a very low percentage of the probability mass and the power of the noise.   
\end{enumerate}
\section{Training in the frequency domain}\label{sec:training_in_the_frequency_domain}
When training in the frequency domain (with the training signal $\xf$) we use only part of the available band for training and leave the rest of the band to transmit data. This section shows the conditions where the lower bound on training energy for almost perfect channel recovery~(\ref{eq:baron_energy}) is achievable despite the reduction in the band allocated for training.\\
The main theorem of this section is based on the 'restricted isometry property' of matrices defined in~\cite{candes_2004_a,candes_2004_b} 
and on the fact that the compressing matrix $F$ (see Section~\ref{sec:channel_model}) whose rows are the $m$ harmonic vectors composing $\xf$ obeys with very high probability~\cite{rudelson_2008}~\cite{vershynin_2010} the restricted isometry property for $2\mathcal{L}(w)$-sparse vectors with as small parameter as we wish, if the number of rows of $F$ obeys:
\begin{eqnarray}
m&\geq &O\left(\mathcal{L}(w)\log k_c\log^4 \mathcal{L}(w)\right)\nonumber\\
&=&O\left(\mathcal{R}_{\hc}\left(\eta_0 \right)\log^3 \mathcal{L}(w)\right)\label{eq:rip_harmonic}
\end{eqnarray}
where the equality (\ref{eq:rip_harmonic}) is based on explicit evaluation of $\mathcal{R}_{\hc}\left(\eta_0 \right)$ in~(\ref{eq:rdf_eta0}).

Recall that $\SNRf$ is the signal to noise ratio of the $m$ channel measurements. The following theorem shows when the channel measurements and the training energy can be minimized together. 
\begin{Theorem}\label{th:5}
If the total training energy is at least $(1+\epsilon)k_c\SNR_0$ (i.e.~$\SNRf\geq (1+\epsilon)\frac{k_c}{m}\SNR_0$) and $m$, the number of harmonic vectors composing the training signal $\xf$ obeys~(\ref{eq:rip_harmonic}),
 then the mean square error of $\hc$ is o(1) in the wideband limit. On the other hand, if the total training energy is less than $(1-\epsilon)k_c\SNR_0$ (i.e.~$\SNRf\leq (1-\epsilon)\frac{k_c}{m}\SNR_0$) then the mean square error of $\hc$ is $1-o(1)$.
\end{Theorem}
\textbf{Interpretation:}
As long as the training signal $\xf$~(\ref{eq:xf}) is composed of enough harmonic vectors, such that the corresponding matrix $F$ obeys the restricted isometry property with a very low parameter, the performance of training is asymptotically the same as training by impulse probing with the same total amount of energy.
Equation~(\ref{eq:rip_harmonic}) shows that if the number of channel measurements is in order of magnitude of the rate distortion function $\mathcal{R}_{\hc}(\eta_0)$  
multiplied by $\log^3 \mathcal{L}(w)$, then recovery is possible using minimum training energy~(\ref{eq:baron_energy}). Can we reduce the number of measurements further and still achieve minimum training energy? 
By~\cite{vershynin_2010} it is known that if the compressing matrix was i.i.d.~Gaussian, the condition on $m$ is
 \begin{equation}\label{eq:rip_gaussian}
m>>\mathcal{R}_{\hc}(\eta_0)
\end{equation}
so for an i.i.d.~gaussian matrix the only condition required to achieve minimum training energy is that $m$ is a superlinear function in $\mathcal{R}_{\hc}(\eta_0)$. In the case of $F$, where the rows of $\h$ are harmonic vectors, we don't know whether the condition~(\ref{eq:rip_harmonic}) can be improved. \\
Using Theorem~\ref{th:5}, training in the frequency domain yields a corollary similar to Corollary~\ref{th:col1} and a theorem and corollary similar to Theorem~\ref{th:2} and Corollary~\ref{th:col2} while using the same total amount of training energy over $\hg$.
\section{Summary}
This paper evaluated the performance of training over $\hg$ and $\hc$ in the low SNR regime. Training over $\hc$ achieves the lower bound on training energy for almost perfect recovery. Moreover, recovery using minimal training energy is possible even using much fewer measurements that the length of the sparse vector. While training with an energy even slightly below $k_c\SNR_0$, the minimum mean square error does not decrease at all, but the penalty term and the rate distortion function are strongly affected. 
\bibliography{refs}
\end{document}